# Hawking radiation from a semi-classical Schwarzschild black hole

**Moslem Shafiee**[1,a], **Ahmad Sheykhi**[1,2,b]

[1] Department of Physics, College of Science, Shiraz University, Shiraz 71454, Iran
[2] Biruni Observatory, College of Science, Shiraz University, Shiraz 71454, Iran

Received: 8 July 2026 / Accepted: 5 September 2026

**Abstract** This study investigates the evaporation process of a Schwarzschild black hole, incorporating quantum corrections arising from conformal anomaly and vacuum polarization. We show that these quantum corrections significantly modify the Hawking radiation and thermodynamic behavior of the black hole. In particular, the Hawking temperature reaches a maximum value and subsequently decreases as the black hole loses mass, eventually vanishing in the extremal limit. The evaporation process therefore approaches an extremal remnant configuration with zero Hawking temperature. Furthermore, we demonstrate that the black hole entropy receives logarithmic corrections due to the modified thermodynamic structure. Hawking radiation is also analyzed using the Parikh–Wilczek tunneling method, which confirms the consistency of the obtained thermodynamic behavior within the quantum-corrected framework.

## 1 Introduction

The pioneering works of Bekenstein and Hawking established that black holes are not purely classical objects, but rather thermodynamic systems characterized by a temperature and an entropy proportional to the area of the event horizon [1–3]. In particular, Hawking's discovery of black hole radiation demonstrated that quantum field theory in curved spacetime leads to thermal particle creation in the vicinity of the event horizon, giving rise to black hole evaporation [2,3]. This profound connection between gravity, thermodynamics, and quantum theory has played a central role in modern theoretical physics and has led to deep conceptual questions regarding the microscopic origin of black hole entropy and the fate of information during evaporation [4–6].

While Hawking's framework provides a robust description for the evaporation of macroscopic black holes, it is expected to encounter significant limitations near the Planck scale, where quantum gravitational corrections to the spacetime geometry become increasingly relevant near the endpoints of the evaporation process. In this regime, purely classical descriptions of spacetime are expected to become insufficient, motivating effective semi-classical approaches in which quantum corrections are incorporated into the background geometry. Such considerations have motivated extensive studies of quantum-corrected and regular black hole geometries, including non-singular black holes with inner horizons [7,8], loop quantum gravity inspired Planck stars [9,10], asymptotically safe black holes [11,12], and models based on generalized uncertainty principles or noncommutative geometry [13–16]. These approaches often predict the existence of black hole remnants and significant deviations from the standard Hawking evaporation scenario. While some of these models introduce modifications through phenomenological parameters, others attempt to incorporate quantum effects through effective descriptions motivated by quantum field theory or quantum gravity considerations.

An important feature shared by many quantum gravity inspired models is the appearance of corrections to the Bekenstein–Hawking entropy formula. In particular, logarithmic corrections to black hole entropy arise in a wide variety of approaches, including loop quantum gravity, quantum geometry, generalized uncertainty principle frameworks, and quantum field theoretic treatments of black hole thermodynamics [17–22]. Such corrections are often interpreted as carrying information about the underlying microscopic structure of quantum spacetime and may become especially important in the final stages of black hole evaporation. Recent investigations have further emphasized the role of logarithmic entropy corrections and quantum modifications in determining the thermodynamic stability and endpoint of evaporating black holes [23–25].

[a] e-mail: m.shafiee72@sharif.edu (corresponding author)
[b] e-mail: asheykhi@shirazu.ac.ir

Published online: 23 September 2026

In a previous work [26], we showed that vacuum quantum effects, as incorporated through leading-order corrections to the Einstein field equations, modify the Schwarzschild geometry in such a way that the classical event horizon splits into two distinct horizons: an outer horizon, which reduces to the classical Schwarzschild horizon in the appropriate limit, and a purely quantum inner horizon. Quantum vacuum effects can modify the late stages of gravitational collapse and may provide a mechanism for avoiding singular behavior within the effective description, thereby allowing for the possibility of non-singular black holes. In such scenarios, after the formation of outer and inner horizons, the collapse is prevented from reaching a singular state and instead undergoes a bounce, initiating a subsequent phase of expansion. From the perspective of an external observer, the timescale associated with this bounce and the return to the inner horizon radius would be extremely long. Consequently, the resulting non-singular black hole would appear effectively stable to distant observers. This modified geometry provides a useful framework for investigating the influence of quantum corrections on Hawking radiation and leads to a richer thermodynamic structure. In particular, these leading-order quantum corrections modify the Hawking temperature, removing the classical divergence and resulting in a finite temperature that gradually approaches zero in the Planck regime. In the present work, we study the Hawking evaporation and thermodynamic properties of this quantum-corrected Schwarzschild black hole. Using both the surface gravity formalism and the Parikh–Wilczek tunneling method, we derive the quantum-corrected Hawking temperature and entropy, analyze the emergence of logarithmic entropy corrections, and investigate the existence of a remnant configuration associated with the final stage of evaporation. In Sect. 2, the structure of the quantum-corrected black hole under consideration is examined. The form of the quantum stress–energy tensor and its components are reviewed, with particular attention to the fact that this tensor consists of distinct contributions representing different quantum effects, including the conformal anomaly and vacuum polarization. By incorporating this quantum stress–energy tensor into the Einstein field equations, the first order quantum-corrected Schwarzschild solution is then introduced. In this framework, the Schwarzschild black hole is found to possess two horizons, namely an outer horizon and an inner horizon. Finally, the modifications induced by quantum corrections on circular photon orbits are discussed as a possible criterion for assessing the observational detectability of these corrections through astrophysical observations. Section 3 is devoted to the study of Hawking radiation in the presence of quantum effects. By calculating the radiation temperature and comparing it with the classical case, we observe that quantum corrections lead to a significant qualitative departure from the classical scenario, fundamentally modifying the thermal behavior of black holes, particularly during the final stages of the evaporation process. In this regime, the black hole undergoes a phase transition, and an analysis of the heat capacity indicates the possibility of the black hole reaching thermal equilibrium with its surrounding environment. Furthermore, the Bekenstein–Hawking entropy acquires a logarithmic correction and becomes dependent not only on the black hole mass but also on parameters associated with quantum fields, which may be interpreted as a form of quantum hair. In Sect. 4, the Parikh–Wilczek tunneling method is employed to investigate Hawking radiation. In this approach [27], conservation of energy is taken into account, so that with the emission of each quantum of radiation, the black hole mass changes accordingly. As a result, the spacetime geometry becomes dynamical, implying that Hawking radiation is no longer exactly thermal. The analysis of the black hole temperature and entropy within this framework reproduces and confirms the results obtained in Sect. 3, thereby demonstrating the internal consistency of the framework established by quantum effects. Throughout this work, we adopt natural units such that $G = c = k_B = \hbar = 1$, which implies that the Planck mass and Planck length are set to unity, i.e., $m_p = l_p = 1$. Nevertheless, in order to make the scale of quantum effects explicit, factors of $m_p$ are retained in all relevant expressions.

## 2 A review on the quantum-corrected Schwarzschild solution

In this section, we review and discuss how the backreaction of quantum fields can modify the structure of the Schwarzschild black hole. For further details and more technical issues, the reader is referred to [26].

The quantum theoretical prediction that the vacuum state possesses a non-zero energy provides strong motivation to reconsider classical solutions of Einstein's equations. When the backreaction of quantum fields is taken into account, the Einstein equations are accordingly reformulated as

$$G_{\mu\nu} = 8\pi \langle T_{\mu\nu} \rangle, \tag{1}$$

where $\langle T_{\mu\nu} \rangle$ denotes the expectation value of the stress–energy tensor of the quantum fields in the vacuum state. The appropriate choice of vacuum depends on the specific spacetime solution under consideration. For instance, if black holes are regarded as the end states of gravitational collapse, then the relevant vacuum state for evaluating the expectation value is the Unruh vacuum. In the classical case, a Schwarzschild black hole is described as a vacuum, spherically symmetric solution of Einstein's field equations, with the metric given

by

$$ds^2 = -\left(1 - \frac{2M}{r}\right)dt^2 + \left(1 - \frac{2M}{r}\right)^{-1}dr^2 + r^2(d\theta^2 + \sin^2\theta d\phi^2), \quad (2)$$

where $M$ is the mass of the black hole.

Imposing spherical symmetry, the vacuum expectation value of the quantum stress–energy tensor possesses four non-vanishing components: an energy density $\rho = -\langle T_t^t \rangle$, an energy flux density $l = -\langle T_t^r \rangle/(1 - 2M/r)$, a radial stress $p_\parallel = \langle T_r^r \rangle$, and a transverse stress $p_\perp = \langle T_\theta^\theta \rangle = \langle T_\phi^\phi \rangle$. These components are not independent, but are related through the local conservation equation of energy–momentum, i.e., $\nabla_\nu \langle T^{\mu\nu} \rangle = 0$. In general, the quantum stress–energy tensor can be decomposed into two contributions [28–31]: a trace part, which is determined by the conformal anomaly, and a regular part that is free of trace anomaly effects. These two parts have fundamentally different physical origins; consequently, each separately satisfies the local energy–momentum conservation law, i.e., $\nabla_\nu \langle T^{\mu\nu(ca)} \rangle = \nabla_\nu \langle T^{\mu\nu(reg)} \rangle = 0$. The regular part of the quantum stress–energy tensor can further be split into two pieces: a diagonal, traceless part associated with the vacuum polarization, and a non-diagonal section that encodes Hawking radiation.

Since massive fields contribute far less than massless fields to the expectation value of the stress–energy tensor, their effects may be safely neglected. Under this assumption, the total contribution to the conformal anomaly from various massless fields can be expressed as [32–36]

$$\langle T_\mu^\mu \rangle = \frac{m_p^2}{2880\pi^2} \sum q_s \left( C_{\mu\nu\rho\sigma} C^{\mu\nu\rho\sigma} + R_{\mu\nu} R^{\mu\nu} - \frac{1}{3} R^2 \right), \quad (3)$$

where $C_{\mu\nu\rho\sigma}$, $R_{\mu\nu}$, and $R$ denote the Weyl tensor, the Ricci tensor, and the Ricci scalar, respectively. The coefficient $q_s$, within the summation over $s$, is a constant determined by the field content and depends solely on the spin of the fields. Its value is given by 1, −13, and 212 for $s = 0, 1, 2$, respectively. In the Schwarzschild spacetime, we have $C_{\mu\nu\rho\sigma} C^{\mu\nu\rho\sigma} = 48M^2/r^6$ and $R_{\mu\nu} = R = 0$. Therefore, the relation of (3) reduces to

$$\langle T_\mu^\mu \rangle = \frac{m_p^2}{60\pi^2} \frac{M^2}{r^6} \sum q_s. \quad (4)$$

Due to the symmetries of the Schwarzschild spacetime, such as spherical symmetry and invariance under radial boosts, constraints are imposed on the conformal anomaly [29–31]. In particular, the invariance of the Schwarzschild geometry under radial boosts implies that $\rho^{ca} = -p_\parallel^{ca}$. The spherical symmetry also gives $\langle T_\theta^{\theta(ca)} \rangle = \langle T_\phi^{\phi(ca)} \rangle = p_\perp^{ca}$. Inserting these conditions into the local energy–momentum conservation relation, one finds

$$p_\parallel^{ca} - p_\perp^{ca} + \frac{r}{2} \partial_r p_\parallel^{ca} = 0. \quad (5)$$

Using (4) and (5), the components of the quantum stress–energy tensor associated with the conformal anomaly can be determined as

$$\rho^{ca} = -p_\parallel^{ca} = \frac{m_p^2}{120\pi^2} \frac{M^2}{r^6} \sum q_s, \quad (6a)$$

$$p_\perp^{ca} = \frac{m_p^2}{60\pi^2} \frac{M^2}{r^6} \sum q_s. \quad (6b)$$

The vacuum polarization contributions to the quantum stress–energy tensor in the Unruh vacuum for the Schwarzschild spacetime, up to first order in $\hbar$, are given by [37–42]

$$\begin{aligned}\rho^{vp} = -p_\parallel^{vp} = p_\perp^{vp} &= \frac{m_p^2}{7680\pi^2} \frac{D}{r^4} \sum A_s \\ &+ \frac{m_p^2}{1280\pi^2} \frac{M}{r^5} \sum A_s (B_s - 1) \\ &- \frac{m_p^2}{768\pi^2} \frac{M^2}{r^6} \sum A_s B_s. \end{aligned} \quad (7)$$

Here, $D$ is an integration constant with the value of 0.621. The coefficients $A_s$ and $B_s$ depend on the spin of the quantum fields; for $s = 0, 1, 2$, they take the respective values of (14.26, 6.49, 0.74) and (0.54, 3.8, 25).

Therefore, by employing Eqs. (6) and (7), the full components of the quantum stress–energy tensor are obtained as

$$\begin{aligned}\langle T_t^t \rangle &= -\rho = -(\rho^{ca} + \rho^{vp}), \\ \langle T_r^r \rangle &= p_\parallel = p_\parallel^{ca} + p_\parallel^{vp}, \\ \langle T_\theta^\theta \rangle &= \langle T_\phi^\phi \rangle = p_\perp = p_\perp^{ca} + p_\perp^{vp}. \end{aligned} \quad (8)$$

Now, substituting the general form of a spherically symmetric metric,

$$ds^2 = -f(r)dt^2 + g(r)dr^2 + r^2(d\theta^2 + \sin^2\theta d\phi^2), \quad (9)$$

into (1), and using the above determined quantum stress–energy tensor, one can show that the quantum-corrected Schwarzschild metric, up to the first order in $\hbar$, takes the form [26]

$$ds^2 = -f(r)dt^2 + f^{-1}(r)dr^2 + r^2(d\theta^2 + \sin^2\theta d\phi^2), \quad (10)$$

where

$$f(r) = \left(1 - \frac{2M}{r} + \frac{cm_p^2}{r^2} + \frac{c' m_p^2 M}{r^3} + \frac{c'' m_p^2 M^2}{r^4}\right), \quad (11)$$

with definitions

$$c = \frac{D\sum A_s}{960\pi} = 0.004,$$
$$c' = \frac{\sum A_s(B_s-1)}{320\pi} = 0.029,$$
$$c'' = \frac{32\sum q_s - 5\sum A_s B_s}{1440\pi} = 1.358. \quad (12)$$

In Eq. (11), the origin of the first and second correction terms in $f(r)$ is exclusively vacuum polarization, whereas the third correction term is jointly due to the conformal anomaly and vacuum polarization. An analysis of the metric function $f(r)$ reveals that it possesses two distinct positive real roots, indicating the presence of two event horizons in the quantum-corrected Schwarzschild spacetime. The larger root, corresponding to the outer horizon, is given by

$$r_+ \simeq \frac{2M}{1+a\left(\frac{m_p}{M}\right)^2}, \quad a = \frac{c}{4} + \frac{c'}{8} + \frac{c''}{16}. \quad (13)$$

This corresponds to the classical Schwarzschild event horizon, which is modified by quantum corrections. The second root, corresponding to the inner horizon, emerges entirely due to quantum backreaction effects and is expressed as

$$r_- \simeq \frac{\left(\frac{c''}{2}\frac{M}{m_p}\right)^{\frac{1}{3}}}{1 - b\left(\frac{c''}{2}\frac{M}{m_p}\right)^{\frac{1}{3}}\frac{m_p}{M}} l_p, \quad b = \frac{2c'+c''}{3c''}. \quad (14)$$

It should be emphasized that, for an astrophysical black hole, the inner horizon is located at a radius far exceeding the Planck length. Also, the trapped region between the inner and outer horizons may attain a considerable spatial extension, as implied by $r_+/r_- \sim M^{\frac{2}{3}}$. Therefore, quantum vacuum effects deform the Schwarzschild spacetime in such a way that the geometry no longer exhibits a lone event horizon; instead, it features two horizons: an outer horizon and a distinct inner horizon.

The direct observation of quantum corrections to the Schwarzschild geometry presents a significant challenge, potentially even proving impossible. This difficulty arises from the extremely minute magnitude of these corrections for genuine astrophysical black holes. The analysis of circular photon orbits serves as a pertinent illustration of this phenomenon. Under the quantum-corrected metric of (10), the condition for the existence of circular photon orbits yields

$$0 = -f(r)dt^2 + r^2(d\theta^2 + \sin^2\theta d\phi^2). \quad (15)$$

Due to the spherical symmetry, all the circular orbits in a constant distance from the center are equivalent, so considering $\theta = constant$ leads to

$$\left(\frac{d\phi}{dt}\right)^2 = \frac{f(r)}{r^2\sin^2\theta}. \quad (16)$$

On the other hand, using the geodesic equation for the $r$ component gives

$$\Gamma^r_{\mu\nu}\frac{dx^\mu}{dt}\frac{dx^\nu}{dt} = 0, \quad \frac{dr}{dt} = \frac{d^2r}{dt^2} = \frac{d\theta}{dt} = 0, \quad (17)$$

which results in

$$\left(\frac{d\phi}{dt}\right)^2 = \frac{f'(r)}{2r\sin^2\theta}, \quad (18)$$

where the prime symbol denotes the derivative with respect to $r$. From the comparison of (16) and (18), we get

$$2f = rf'. \quad (19)$$

By solving the above equation, the value of the radial distance for the circular orbit is determined as

$$r_{co} \simeq \frac{3M}{1+d\left(\frac{m_p}{M}\right)^2} = 3M\left(1 - d\left(\frac{m_p}{M}\right)^2 + \mathcal{O}\left(\frac{m_p^4}{M^4}\right)\right), \quad (20)$$

where $d = \frac{2}{9}c + \frac{5}{54}c' + \frac{1}{27}c''$. Classically, the innermost stable circular orbit for photons resides at $r_{isco} = 1.5R_s$, where $R_s$ is the Schwarzschild radius. Considering that for a solar-mass black hole, the quantum parameter $(m_p/M)^2 \approx 10^{-76}$, the impact of quantum corrections on the spacetime geometry and the resulting circular orbits is negligible. Consequently, distinguishing these quantum-affected orbits from their classical counterparts via astrophysical observations is highly problematic.

In the following section, we will discuss how the quantum corrections to the Schwarzschild geometry affect Hawking radiation and the associated thermodynamic properties.

## 3 Quantum corrections to Hawking temperature and black hole entropy

In the classical picture, when a black hole begins to radiate, its mass gradually decreases. As Hawking radiation continues and the black hole shrinks, the evaporation process accelerates, eventually leading to the complete disappearance of the black hole. However, this scenario is modified once quantum vacuum effects are taken into account.

The Hawking temperature of a black hole is given by the following relation [2,3]

$$T_H = \frac{\kappa}{2\pi}, \quad (21)$$

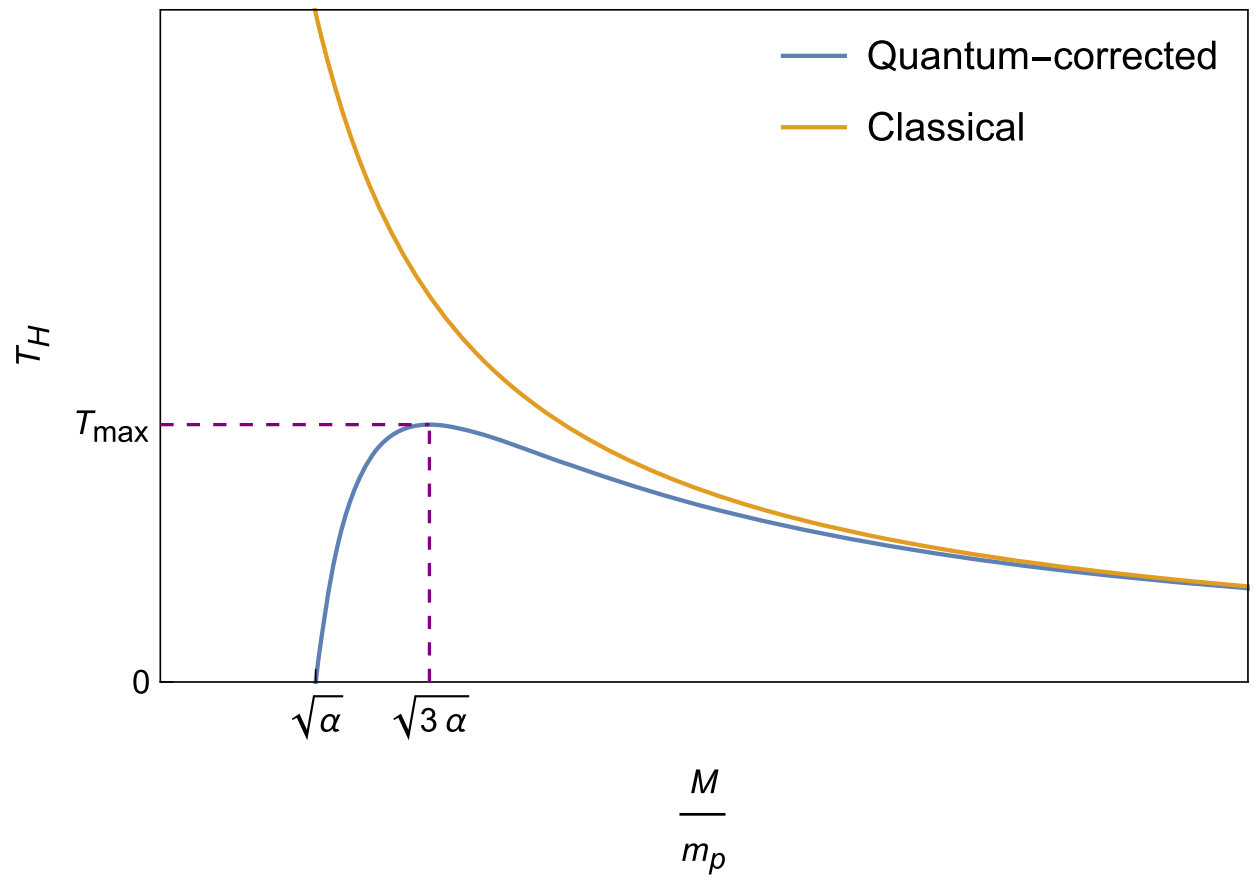

**Fig. 1** A schematic plot of the Hawking temperature in the classical and semi-classical regimes. Contrary to the classical case, quantum corrections cause the black hole temperature to be maximal when the black hole mass reaches the critical value of $M_c = \sqrt{3\alpha} m_p$. Subsequently, the Hawking temperature begins to decrease, such that eventually, at $M_{min} = \sqrt{\alpha} m_p$, the temperature vanishes and the evaporation process ceases

where $\kappa$ denotes the surface gravity at the event horizon. For a metric of the form (10) with $f(r)$ given by (11), the temperature can be expressed as

$$T_H = \frac{1}{4\pi}\left(\frac{df}{dr}\right)\bigg|_{r=r_+}. \tag{22}$$

Computing this up to first order in $\hbar$, we obtain

$$\begin{aligned} T_H &= \frac{1}{8\pi M} - \frac{c' + c''}{64\pi}\frac{m_p^2}{M^3} \\ &= \frac{1}{8\pi M}\left(1 - \left(\frac{M_{min}}{M}\right)^2\right), \end{aligned} \tag{23}$$

where in the last equality the following definitions have been used

$$\alpha = \frac{c' + c''}{8}, \quad M_{min} = \sqrt{\alpha} m_p. \tag{24}$$

It is observed that incorporating quantum vacuum effects leads to a reduction of the Hawking temperature compared to the classical case. As the evaporation proceeds, the temperature of the black hole increases until, at the critical mass $M_c = \sqrt{3\alpha} m_p$, it reaches its maximum value, ie,

$$T_{max} = \frac{1}{12\pi M_c} = \frac{1}{12\pi\sqrt{3\alpha} m_p}. \tag{25}$$

At this stage, the temperature begins to decrease and eventually, when the mass reaches $M_{min} = \sqrt{\alpha} m_p$, the Hawking temperature drops to zero and the evaporation process comes to a halt. Figure 1 illustrates the behavior of the Hawking temperature under quantum corrections and highlights its deviation from the classical case.

The timescale of evaporation, as the temperature begins to decrease, becomes increasingly prolonged. Utilizing the Stefan–Boltzmann law and the relation of (23) for temperature, it can be demonstrated that in the final stages of evaporation, we have $t \sim (M - M_{min})^{-3}$. In other words, the time required for the radiation to cease will be exceedingly long. Therefore, for a Schwarzschild black hole with quantum corrections, although the radiation does not continue to complete evaporation, the time taken to reach zero temperature and halt the evaporation process will be significantly longer than that for the complete evaporation of a classical Schwarzschild black hole.

The values of $M_{min}$, $M_c$, and consequently $T_{max}$, depend on the parameter $\alpha$. Here, $\alpha$ is an effective quantum parameter whose value is determined by the properties and the content of the quantum fields. For example, for massless bosonic fields with spins 0, 1, and 2, one finds $\alpha = 0.17$; hence $M_{min} = 0.4 m_p$ and $M_c = 0.7 m_p$. Therefore, irrespective of the numerical value of $\alpha$, one expects that, as the black hole mass approaches the Planck scale, the Hawking radiation exhibits a qualitatively different behavior.

When the black hole begins to radiate and its mass decreases, the outer and inner horizons approach each other, and the trapped region between them becomes smaller. When the Hawking temperature goes to zero, the two outer and inner horizons coincide and the trapped region disappears. In Fig. 2, the evolution of the outer and inner horizons is illustrated as the Schwarzschild black hole mass decreases due to Hawking radiation. Therefore, the final state produced by the evaporation of a Schwarzschild black hole with quantum corrections will be an extremal remnant. Compared with the extremal Reissner–Nordström black hole, in view of the absence of electric charge, the remnant that forms will be stable.

Using the first law of black hole thermodynamics ($dM = T dS_{BH}$), one can compute the black hole entropy. With the help of (23), the entropy is obtained to first order of $\hbar$ as

$$S_{BH} = 4\pi M^2 + 8\pi\alpha m_p^2 \ln M. \tag{26}$$

The first term is precisely the Bekenstein–Hawking entropy for classical Schwarzschild black holes, while the second term has a logarithmic form and originates from quantum corrections. Therefore, the entropy depends not only on the black hole mass, but also on the type and nature of the quantum fields considered. As indicated by (26), for $M > m_p$ the entropy is larger than the classical value, for $M = m_p$ it coincides exactly with the classical entropy, and for $M_{min} \le M < m_p$, it becomes smaller than the classical result.

Thus, one may conclude that quantum corrections effectively increase the number of degrees of freedom; however, as the evaporation starts and the black hole mass decreases, the entropy approaches the classical value, and eventually

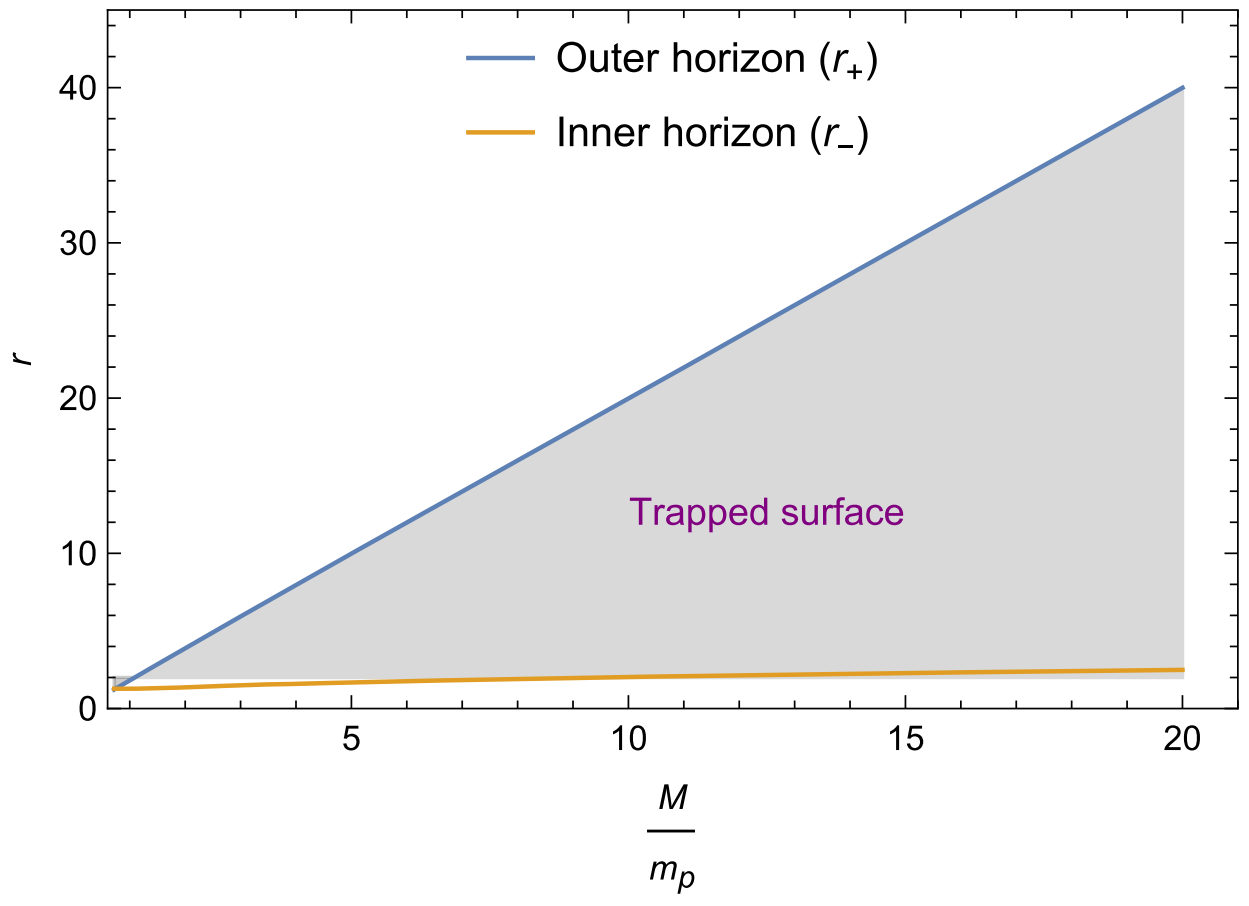

**Fig. 2** The plot of the outer and inner horizons as a function of the black hole mass, taking into account the numerical values for massless bosonic fields with spins 0, 1, and 2. Hawking radiation drives the two horizons to approach each other and shrinks the trapped surface between them (the region shown between the two curves). In the final stage of evaporation, the two horizons merge and the trapped region disappears, thereby resulting in the formation of an extremal remnant

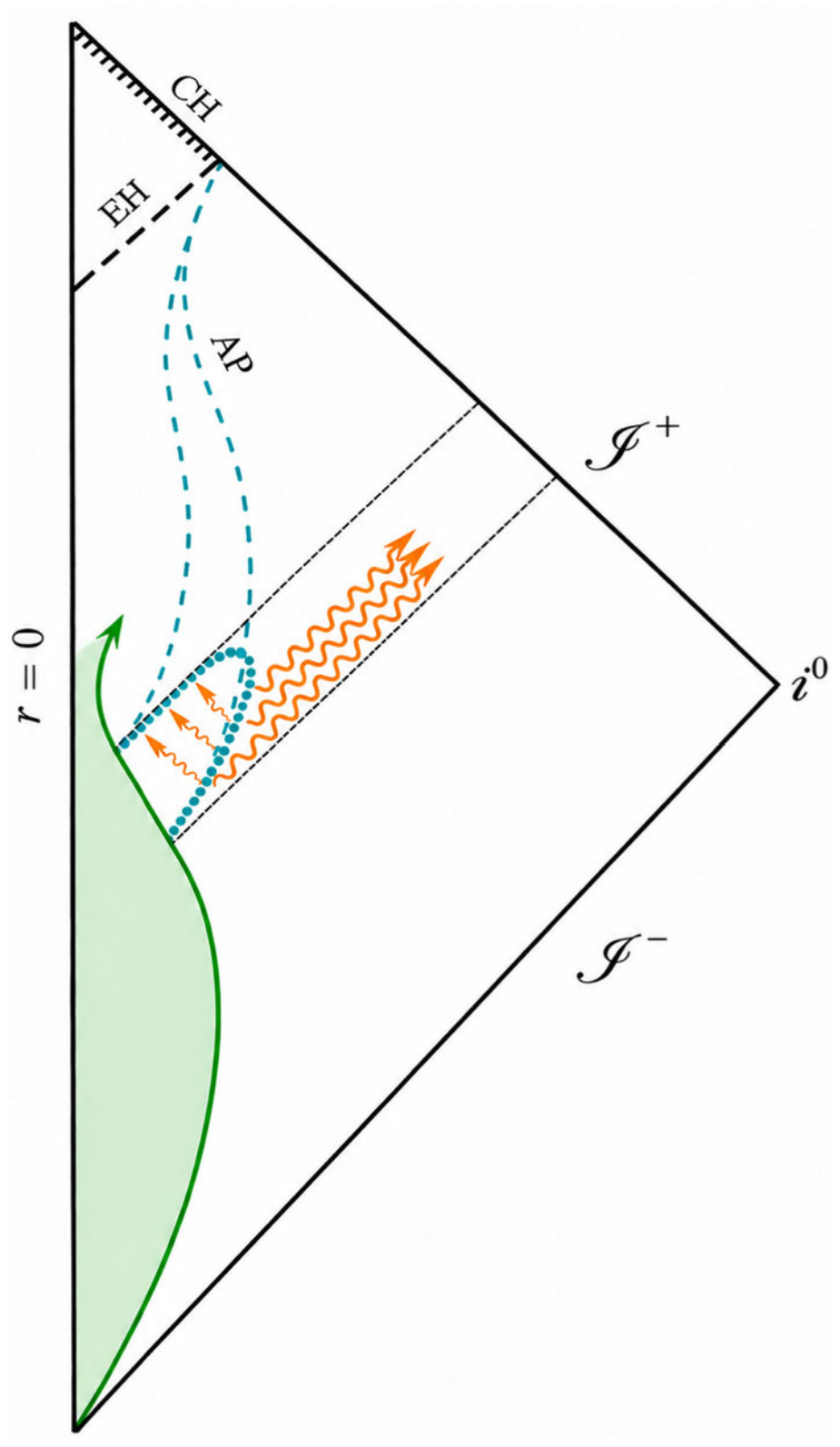

**Fig. 3** The Carter–Penrose diagram of a quantum-corrected Schwarzschild black hole formed through gravitational collapse is shown. The green solid curve represents the surface of the collapsing star. The black dashed lines correspond to the outer and inner horizons. The orange jagged arrows denote Hawking radiation, including both ingoing and outgoing modes. Once evaporation begins, the outer and inner horizons, corresponding to the blue dashed trajectories, progressively converge until they eventually merge into a single horizon, at which point the trapped region disappears. Quantum corrections enable the formation of a non-singular black hole [26]: after the second horizon appears, the collapse stops and the stellar core begins to expand. The final phase of evaporation process is an extremal remnant with vanishing Hawking temperature

within the Planckian regime, it becomes smaller than the classical one. This implies a reduction in the number of available microstates at that scale.

The entropy has been expressed in terms of the black hole mass, it can also be written in terms of the horizon area by using the relation $A = 4\pi r_+^2$. In this case, the entropy becomes

$$S_{BH} = \frac{k_B c^3 A}{4G\hbar} + 4\pi\alpha k_B \ln\left(\frac{c^3 A}{16\pi G\hbar}\right). \tag{27}$$

In this expression, we restore all physical constants that had previously been set to unity, so that the result is presented in explicit physical units rather than in natural units.

The analysis of the black hole heat capacity ($C = \frac{dM}{dT_H}$) shows that in the regime $M > M_c$, similarly to the classical case, the black hole exhibits a negative heat capacity. However, once the black hole temperature reaches its maximal value, for $M < M_c$, the heat capacity becomes positive. This implies that in the near-extremal regime, the black hole can attain thermal equilibrium with the surrounding environment. In other words, when the black hole mass approaches the critical value, i.e., $M_c$, a phase transition occurs such that the black hole evolves from a thermodynamically unstable configuration to a thermodynamically stable state.

In summary, the backreaction of quantum fields modifies the black hole geometry and its thermodynamic behavior, particularly in the final stages of evaporation. Under these conditions, the black hole develops a maximum Hawking temperature; before reaching this stage, Hawking radiation behaves similarly to the classical case. After passing the critical point, however, the Hawking temperature starts to decrease and eventually vanishes. The Hawking radiation drives the inner and outer horizons to approach each other, so that by the end of the process the horizons merge and the trapped surface disappears. The final phase is therefore an extremal remnant with vanishing Hawking temperature, corresponding to a stable configuration. Figure 3 displays the Carter–Penrose diagram illustrating the formation of a quantum-corrected black hole via gravitational collapse and its evolution up to the final stages of evaporation.

Hence, within the framework of the effective semi-classical model considered here, the inclusion of quantum effects leads to significant deviations from the classical behavior, including a modified temperature profile and a slowed evaporation process. Interestingly, these qualitative features are consistent with results obtained in a variety of other quantum-corrected black hole models in the litera-ture [8–15,43–47], suggesting that they may capture generic features of semi-classical backreaction effects rather than model-specific behavior. An important feature of the present model is that the modified geometry is motivated by vacuum quantum corrections to the Schwarzschild spacetime rather than being introduced through an arbitrary phenomenological deformation.

## 4 Tunneling description of Hawking radiation

One may interpret the outgoing radiation from the black hole as a particle tunneling process [27]. In this picture, a positive energy (negative energy) particle from a virtual pair can tunnel outward (inward) from just inside (outside) the outer horizon. In accordance with energy conservation, each emission with energy $\omega$ reduces the mass of the black hole by the same amount, in that, $M-\omega$ [48].

With this consideration, the spacetime geometry becomes dynamical once Hawking radiation is taken into account. Thus, if the radial positions of the positive energy particle before and after tunneling are denoted by $r_{in}$ and $r_{out}$, respectively, then we have

$$r_{in}=r_+(M)-\epsilon=\frac{2M}{1+a\left(\frac{m_p}{M}\right)^2}-\epsilon, \tag{28a}$$

$$r_{out}=r_+(M-\omega)+\epsilon=\frac{2(M-\omega)}{1+a\left(\frac{m_p}{M-\omega}\right)^2}+\epsilon, \tag{28b}$$

where $\epsilon$ is a very small parameter with the dimension of length. As mentioned earlier, after the emission of radiation and the consequent decrease in the black hole mass, the outer horizon no longer remains at its original position and is shifted.

The tunneling probability of a particle is determined by the imaginary part of its action, in that,

$$Im\ S=Im\int_{r_{in}}^{r_{out}}p_r dr=Im\int_{r_{in}}^{r_{out}}\int_0^{p_r}dp_r' dr, \tag{29}$$

where $p_r$ is the canonical radial momentum of the particle. Using Hamilton's equation $\dot{r}=\frac{\partial H}{\partial p_r}$, where $H=M-\omega'$ is the Hamiltonian, the above integral can be rewritten as

$$Im\ S=Im\int_{r_{in}}^{r_{out}}\int_M^{M-\omega}\frac{dH}{\dot{r}}dr. \tag{30}$$

By interchanging the order of integration, one finds

$$Im\ S=Im\int_0^{\omega}\int_{r_{in}}^{r_{out}}\frac{dr}{\dot{r}}(-d\omega'). \tag{31}$$

To determine $\dot{r}$ and evaluate the above integral, the metric of (10) must be written in a coordinate system that is regular at the outer horizon. For this purpose, by introducing the Painlevé–Gullstrand coordinate transformation of the form

$$dt\to dt-\frac{\sqrt{1-f(r)}}{f(r)}, \tag{32}$$

the metric of (10) takes the form

$$ds^2=-f(r)dt^2+2\sqrt{1-f(r)}dtdr +dr^2+r^2(d\theta^2+\sin^2\theta d\phi^2). \tag{33}$$

Now, for a massless particle and assuming radial emission, we obtain

$$\dot{r}=1-\sqrt{1-f(r)}. \tag{34}$$

To evaluate the integral of (31), it suffices to obtain the value of $\dot{r}$ in the vicinity of $r_+$, in that,

$$f(r)\simeq f(r_+)+f'(r_+)(r-r_+) \Rightarrow \dot{r}=\frac{1}{2}f'(r_+)(r-r_+), \tag{35}$$

Thus

$$Im\ S=Im\int_0^{\omega}\int_{r_{in}}^{r_{out}}\frac{-2drd\omega}{f'(r_+)(r-r_+)}. \tag{36}$$

This integral possesses a simple pole at $r_+$, which can be readily calculated by deforming the contour. Consequently, we have

$$Im\ S=Im\int_0^{\omega}\frac{\pi d\omega'}{\frac{1}{4(M-\omega')}\left(1-\alpha\frac{m_p^2}{(M-\omega')^2}\right)} =Im\int_0^{\omega}4\pi(M-\omega')\left(1+\alpha\frac{m_p^2}{(M-\omega')^2}\right)d\omega', \tag{37}$$

which results in

$$Im\ S=4\pi\left(M\omega-\frac{\omega^2}{2}\right)+4\pi\alpha m_p^2\ln\frac{M}{M-\omega}. \tag{38}$$

On the other hand, Hawking radiation can be considered as a tunneling of negative energy particles from outside the black hole into it. In this picture, the black hole mass after the emission will be $M+\omega$. By performing the corresponding calculations in this scenario, we arrive at a result similar to relation of (38).

Given that near the outer horizon, the radial wavenumber of the particle becomes very large, and its wavelength is strongly blueshifted, we can therefore expect the WKB approximation to be valid. In this case, the tunneling probability of the particles becomes approximately equal to [49,50]

$$\Gamma\sim e^{-2Im\ S}=e^{\Delta S_{BH}}, \tag{39}$$

where the factor of 2 appears due to taking into account both tunneling channels of the particle and the antiparticle. Therefore, the corresponding change in entropy becomes

$$\Delta S_{BH} = S_{BH}(M-\omega) - S_{BH}(M) \\ = -8\pi\left(M\omega - \frac{\omega^2}{2}\right) - 8\pi\alpha m_p^2 \ln\frac{M}{M-\omega}. \quad (40)$$

This relation is consistent with the expression obtained for the black hole entropy in (26). In the case of $M >> \omega$, $e^{-2Im\ S}$ transforms into the Boltzmann factor of a particle with energy $\omega$, i.e., $e^{-\frac{\omega}{T}}$. Therefore, by considering terms linear in $\omega$ and neglecting higher-order terms, we obtain

$$\frac{\omega}{T} = 8\pi M\omega\left(1+\alpha\left(\frac{m_p}{M}\right)^2\right), \quad (41)$$

which is in perfect agreement with (23). Thus, in the limit where the emitted energy is small relative to the black hole mass, the temperature is determined by the derived relation, and the radiation appears thermal. Consequently, interpreting Hawking radiation as particle tunneling confirms our previous results.

The higher-order terms in $\omega$, arising from the consideration of energy conservation, become significant as the black hole mass diminishes. This leads to a highly non-thermal radiation spectrum, which could drastically alter the black hole's information-theoretic behavior at such scales.

## 5 Conclusion and discussion

In this work, we have investigated the Hawking radiation and thermodynamic properties of a quantum-corrected Schwarzschild black hole arising from vacuum quantum effects within an effective semi-classical framework. The quantum modified geometry considered here leads to a nontrivial causal structure characterized by the emergence of two distinct horizons: an outer horizon, which approaches the classical Schwarzschild horizon in the appropriate limit, and an inner horizon of purely quantum origin. Such a structure significantly alters the thermodynamic behavior of the black hole, particularly during the late stages of evaporation where quantum effects become increasingly important.

Using both the surface gravity formalism and the Parikh–Wilczek tunneling approach [27,51], we derived the quantum-corrected Hawking temperature associated with the outer horizon of the quantum modified geometry. Our analysis shows that the Hawking temperature reproduces the standard classical behavior for sufficiently large black hole masses, thereby recovering the classical Schwarzschild limit. However, substantial deviations arise near the final stages of evaporation. In particular, unlike the standard Hawking temperature which diverges as the mass decreases, the quantum modified temperature reaches a finite maximum value and subsequently decreases toward zero at a finite critical mass. Black hole evaporation causes the inner and outer horizons to converge, eventually merging into a single horizon in the final stage, which consequently leads to the disappearance of the trapped surface. This behavior indicates the formation of an extremal remnant configuration that halts the evaporation process.

The existence of black hole remnants has been extensively discussed in several approaches to quantum gravity and quantum-corrected black hole physics, including generalized uncertainty principle models [13,43,44], loop quantum gravity inspired geometries [9,10,45], noncommutative black holes [14], regular black holes [8,46,47], and asymptotically safe gravity scenarios [11,12]. Our results exhibit qualitative similarities with many of these approaches, particularly regarding the existence of a maximal temperature, the suppression of complete evaporation, and the emergence of a stable remnant. Nevertheless, an important feature of the present model is that the modified geometry is obtained systematically by incorporating vacuum quantum corrections into the Einstein equations, rather than through phenomenological modifications introduced by hand. Therefore, the resulting quantum-corrected geometry provides an effective framework for investigating how vacuum fluctuations modify black hole evaporation, Hawking radiation, and thermodynamic properties.

It is worth emphasizing that different approaches to semiclassical quantum corrections may lead to different effective modifications of classical black hole geometries. In particular, the renormalized quantum stress–energy tensor depends on several factors, including the choice of quantum state, the background spacetime, and the employed renormalization and computational procedures. As a result, different semi-classical frameworks may produce different correction terms in the effective metric. Such ambiguities have been discussed in detail in studies of vacuum polarization and quantum backreaction effects [29–31]. More recent investigations have also highlighted the dependence of the resulting corrections on the adopted quantum state and background geometry [52–54]. Therefore, differences among various quantum-corrected black hole models should be interpreted as reflecting different realizations of semi-classical quantum effects rather than inconsistencies between the approaches.

We further computed the entropy associated with the quantum-corrected black hole geometry and found that the Bekenstein–Hawking area law acquires logarithmic corrections. Such logarithmic terms are known to appear in a wide variety of quantum gravity inspired frameworks, including loop quantum gravity [17,55], Euclidean quantum gravity [18], quantum geometry approaches [56,57], and generalized uncertainty principle models [19,20]. In the present model, these corrections naturally emerge from the modified thermodynamic structure induced by the quantum-corrected metric.

The resulting entropy therefore supports the idea that logarithmic corrections may reflect a common feature of quantum corrections appearing in different approaches to quantum gravity.

The presence of an inner quantum horizon and the formation of an extremal remnant may also have important implications for the final stages of black hole evaporation and the information loss problem originally raised by Hawking [4,5]. Since the evaporation process terminates at a finite remnant mass rather than ending in a complete disappearance of the black hole, it becomes conceivable that part of the information content may remain stored within the remnant configuration. Although the present work does not attempt to resolve the black hole information paradox, the modified causal structure of the geometry suggests that quantum vacuum effects could play a nontrivial role in the late-time evolution of evaporating black holes.

Moreover, the modified evaporation dynamics found here may influence the Page time and the associated information recovery process [58,59]. Since the quantum-corrected geometry substantially alters the near-extremal phase of evaporation, one may expect corresponding modifications to the Page curve and the entanglement structure of Hawking radiation. In recent years, significant progress has been achieved in understanding black hole information recovery through quantum extremal surfaces and island constructions [60–62]. It would therefore be interesting to investigate whether the quantum-corrected geometry considered in the present work admits similar structures and how the existence of two horizons affects the entanglement entropy of Hawking radiation. A complete investigation of these questions, however, requires a more detailed analysis of quantum information flow and backreaction effects beyond the semi-classical approximation employed here.

Several open problems remain for future investigation. In particular, a fully dynamical treatment of the evaporation process including self-consistent quantum backreaction would be highly desirable. It would also be important to study the stability properties of the remnant configuration under both classical and quantum perturbations. Furthermore, extending the present analysis to rotating and charged black holes may provide deeper insight into the interplay between quantum corrections, horizon structure, and black hole thermodynamics. Such investigations could help clarify the role of quantum vacuum effects in the final stages of gravitational collapse and the microscopic nature of black hole entropy.

Overall, the results obtained in this work suggest that quantum vacuum corrections can substantially modify the late-time behavior of Hawking evaporation and naturally lead to remnant configurations without introducing ad hoc assumptions. The emergence of logarithmic entropy corrections, the existence of an inner quantum horizon, and the modification of the evaporation endpoint collectively indicate that quantum effects may significantly alter the thermodynamic and causal structure of black holes near the Planck scale. We hope that the present work may provide a useful framework for future studies of quantum-corrected black holes, black hole information, and the interface between gravity, thermodynamics, and quantum theory.

**Acknowledgements** We thank the referee for very constructive comments which helped us improve our paper significantly. We are grateful to Shiraz University Research Council. This work is based upon research funded by the Iran National Science Foundation (INSF) and the Ministry of Science under project No. 40401028.

**Data Availability Statement** This manuscript has no associated data. [Authors' comment: Data sharing not applicable to this article as no datasets were generated or analysed during the current study.]

**Code Availability Statement** This manuscript has no associated code/software. [Authors' comment: Code/Software sharing not applicable to this article as no code/software was generated or analysed during the current study.]